\documentclass[useAMS,usenatbib]{mn2e}
\usepackage{graphicx}
\usepackage{aas_macros}
\usepackage{fixltx2e}
\bibpunct{(}{)}{;}{a}{,}{,}

\newcommand\logl{$\log (L/\lsun)$}
\newcommand\logT{$\log (T_{eff}/{\rm K})$}
\newcommand\chisq{$\chi^2~$}
\newcommand\mhz{${\mu}$Hz}
\newcommand\msun{{\rm M}_\odot}
\newcommand\lsun{{\rm L}_\odot}
\newcommand\eg{{\em e.g}\/. }

\newcommand\vsini{$v \sin i$}
\newcommand\DSS{$\delta$~Scuti star}
\newcommand\DS{$\delta$~Scuti}
\title[{\em MOST} observations of the Herbig Ae $\delta$-Scuti star HD~34282]{{\em MOST}\thanks{Based on data from the {\it MOST} satellite, a Canadian Space Agency mission, jointly operated by Dynacon Inc., the University of Toronto Institute for Aerospace Studies and the University of British Columbia with the assistance of the University of Vienna.} observations of the Herbig Ae $\delta$-Scuti star HD~34282.}
\author[M. P. Casey et al.]{M. P. Casey,$^{1}$\thanks{E-mail: mcasey@ap.smu.ca} 
K. Zwintz,$^{2}$
D. B. Guenther,$^{1}$
W. W. Weiss,$^{2}$
P. J. Amado,$^{3}$
\newauthor  D. D\'{i}az-Fraile,$^{3}$
E. Rodriguez,$^{3}$
R. Kuschnig,$^{2}$
J. M. Matthews,$^{4}$ 
A. F. J. Moffat,$^{5}$
\newauthor J. F. Rowe,$^{6}$
S. M. Rucinski,$^{7}$
and D. Sasselov$^{8}$\\
$^{1}$ Institute for Computational Astrophysics, Department of Astronomy and Physics, Saint Mary's University, Halifax, NS~~B3H 3C3, Canada\\
$^{2}$ University of Vienna, Institute of Astronomy, T\"urkenschanzstrasse 17, A-1180 Vienna, Austria \\
$^{3}$ Instituto de Astrof\'{i}sica de Andaluc\'{i}a, CSIC, P.O. Box 3004, 18080, Granada, Spain \\
$^{4}$Department of Physics and Astronomy, University of British Columbia, 6224 Agricultural Road, Vancouver, BC~~V6T 1Z1, Canada \\
$^{5}$D\'epartment de physique, Universit\'e de Montr\'eal, C.P. 6128, Succ. Centre-Ville, Montr\'eal, QC~~H3C 3J7, Canada \\
$^{6}$NASA Ames Research Center, Moffett Field, CA 94035, USA \\
$^{7}$Department of Astronomy and Astrophysics, University of Toronto, 50 St. George St., Toronto, ON~~M5S 3H4, Canada \\
$^{8}$Harvard-Smithsonian Center for Astrophysics, 60 Garden Street, Cambridge, MA 02138, USA 
}

\begin{document}

\date{Accepted 2012 October 16.  Received 2012 October 16; in original form 2012 August 25.}

\pagerange{\pageref{firstpage}--\pageref{lastpage}} \pubyear{2012}

\maketitle

\label{firstpage}

\begin{abstract}
{\em MOST} observations and model analysis of the Herbig Ae star HD~34282 (V1366~Ori) reveal \DS\, pulsations.
22 frequencies are observed, 10 of which confirm those previously identified by \citet{2006MmSAI..77...97A}, and 12 of which  are newly discovered in this work.
We show that the weighted-average frequency in each group fits the radial p-mode frequencies of viable models.
We argue that the observed pulsation spectrum extends just to the edge to the acoustic cut-off frequency and show that this also is consistent with our best-fitting models.
\end{abstract}


\section{HD~34282 peculiarities}
\label{sec:HD34282}

HD~34282 (V1366~Ori, PDS~176) is a Herbig Ae (HAe) star  (the cool A-range subset of the Herbig AeBe stars) that is also a \DSS. Pulsations were originally discovered by \citet{2004MNRAS.352L..11A}, with ten frequencies later identified by \citet{2006MmSAI..77...97A} using multi-site ground-based photometry.
The observed frequencies, from 64.7 to 79.4 cycles/day (d$^{-1}$), are among the highest frequencies detected in a \DSS.
In order to resolve in more detail the star's unusual oscillation spectrum, the {\em Microvariability and Oscillations of STars} ({\em MOST}) satellite \citep{2003PASP..115.1023W} observed HD~34282 for 31~days in 2007 December.
These observations reveal a unique spectrum:  22 frequencies detected, forming groups of frequencies at regular intervals of about 44~\mhz, that may be indicative of the large spacing between successive orders of radial pulsation.
The amplitudes of these groups steadily grow to the highest-frequency group at around 79~d$^{-1}$, above which point there is an abrupt cut off in pulsation power, which we will show is consistent with the star pulsating just below the acoustic cut-off frequency.  Importantly, the  acoustic cut-off frequency has not been previously identified as a factor in the pulsation spectrum of  any \DSS.
Here we report on the results of the {\em MOST} observations and our attempts to model the oscillation frequencies that have led to these conclusions.

HD~34282 was first identified as a HAe object by \citet{1994A&AS..104..315T}, and is therefore assumed to be a pre-main-sequence (PMS) star.
Spectral classifications have ranged from  A0 to A3 \citep[\eg][]{2003AJ....126.2971V, 2001A&A...378..116M, 2004A&A...419..301M}.
The latest Hipparcos reductions report a parallax of 5.2~$\pm$~1.7 milliarcseconds, corresponding to a distance of 191~$^{+89}_{-46}$~pc \citep{2007A&A...474..653V}.
The above spectral classifications place HD~34282 blueward of the classical instability strip, relatively close to the zero-age main sequence (ZAMS; see Fig.~\ref{fig:HRDpos}).
At this position in the Hertzsprung-Russell (HR) diagram, the star's PMS nature is ambiguous - it has either already arrived, or is just about to arrive on the ZAMS.

\citet{2001A&A...378..116M} report a projected rotational velocity (\vsini)$~= 129 \pm 11$~km~s$^{-1}$.
\citet{2004A&A...419..301M} found that HD~34282 has an anomalously low metallicity, [Fe/H] = -0.8 (fractional metal content by mass, Z=0.004).
Such a low metallicity for a supposedly PMS star in the disk of the Milky Way poses questions regarding the true evolutionary status of the star;  unless there are patches of low-metallicity material within the interstellar medium from which the star could form, a newly-formed star should not otherwise have such low metal content.
One solution is that perhaps this star is also a $\lambda$-Bootis star, with depressed levels of heavier metals at the surface, but near-solar values for carbon, nitrogen, and oxygen (CNO) throughout, and near-solar values for heavy metals below the surface layers of the star as well.
The $\lambda$-Bootis characteristics could be the result of recent preferential accretion of metal-depleted gas over metal-rich dust, \eg~as outlined in \citet{2002MNRAS.335L..45K}.
In this case HD~34282 could consistently be a PMS (or new main-sequence) star without otherwise needing to question the low metal abundance of the star.
In the past, \citet{1998AJ....116.2530G} observed this star as part of a systematic campaign to identify $\lambda$-Bootis characteristics in young A-type stars, in which they classified HD~34282 as  {\em A0.5~Vb~(shell)r} in their extended Morgan-Keenan classification system, but failed to detect $\lambda$-Bootis characteristics.
On the other hand \citet{1998AJ....116.2530G} did {\em not} report an anomalously low Z for the star, as later found by 
\citet{2004A&A...419..301M}.
The source of this discrepancy is unknown.
Low metallicity and high \vsini~are confirmed by Amado et al. (in preparation) for heavier metals, but analysis of the required lighter metals (CNO plus S), needed to determine $\lambda$~Bootis status, have not yielded an answer one way or the other -- blending of the relevant spectral lines with other lines caused by the high \vsini~of HD~34282 is beyond the current capabilities of stellar atmosphere models.

We now present the {\em MOST} observations and data reductions of this star (Section~\ref{sec:MOST}),  refine the position of HD~34282 in the HRD using Tycho data (Section~\ref{sec:HRD_pos}), and show the results of an asteroseismic analysis of the {\em MOST} observed frequencies (Section~\ref{sec:asteroseis}).

\section{{\em MOST} observations}
\label{sec:MOST}
\begin{table}
\caption{Pulsation frequencies in order of decreasing SigSpec significance (sig column) for HD~34282 in d$^{-1}$ and $\mu$Hz, with respective last-digit errors given in parentheses.   Amplitudes are in mmag, S/N values were derived from Period04.  The A06 ID cross references the same frequencies as identified by \citet{2006MmSAI..77...97A}.}
\label{tab:freqs}
\begin{scriptsize}
\begin{tabular}{lcccccc}
\hline
 ID  & freq & freq & amp & sig & S/N & A06 \\
     & [d$^{-1}$] & [$\mu$Hz] & [mmag]  &   &   & ID\\
 \hline
$f_1$	&	79.423(1)	&	919.24(1)	&	6.344	&	945.14	&	45.50	&	A10 \\
$f_2$	&	79.252(2)	&	917.27(2)	&	3.523	&	378.97	&	47.61	&	A9 \\
$f_3$	&	75.416(2)	&	872.87(2)	&	3.339	&	360.40	&	23.07	&	A7 \\
$f_4$	&	75.864(2)	&	878.05(3)	&	2.427	&	227.04	&	22.42	&	A8 \\
$f_5$ 	&	75.356(2)	&	872.18(3)	&	2.205	&	189.99	&	31.13	&	A6 \\
$f_6$	&	71.589(3)	&	828.58(3)	&	2.075	&	171.87	&	12.37	&	A4 \\
$f_7$	&	71.525(3)	&	827.84(3)	&	1.862	&	158.92	&	10.07	&	A3 \\
$f_8$	&	57.060(3)	&	660.42(3)	&	1.681	&	150.24	&	17.38	&	A1 \\
$f_9$	&	71.972(2)	&	833.01(3)	&	1.630	&	183.24	&	14.77	&	A5 \\
$f_{10}$	&	68.152(3)	&	788.80(3)	&	1.475	&	147.62	&	17.91	&	A2 \\
$f_{11}$	&	64.695(3)	&	748.79(4)	&	1.245	&	119.38	&	20.16	&	- \\
$f_{12}$	&	61.053(4)	&	706.63(5)	&	0.754	&	60.01	&	12.23	&	- \\
$f_{13}$	&	67.787(5)	&	784.58(5)	&	0.714	&	54.56	&	10.37	&	- \\
$f_{14}$	&	71.043(5)	&	822.26(5)	&	0.707	&	50.67	&	7.22	&	- \\
$f_{15}$	&	75.402(4)	&	872.70(5)	&	0.658	&	56.83	&	6.94	&	- \\
$f_{16}$	&	53.427(5)	&	618.37(6)	&	0.553	&	45.81	&	10.96	&	- \\
$f_{17}$	&	67.534(5)	&	781.65(6)	&	0.524	&	48.45	&	9.83	&	- \\
$f_{18}$	&	75.448(6)	&	873.24(7)	&	0.492	&	32.21	&	14.60	&	- \\
$f_{19}$	&	68.669(6)	&	794.78(7)	&	0.448	&	32.34	&	6.84	&	- \\
$f_{20}$	&	72.319(6)	&	837.03(7)	&	0.418	&	28.38	&	5.43	&	- \\
$f_{21}$	&	60.353(6)	&	698.53(7)	&	0.377	&	27.20	&	7.98	&	- \\
$f_{22}$	&	67.465(7)	&	780.84(8)	&	0.345	&	22.67	&	7.05	&	- \\
\hline
\end{tabular}
\end{scriptsize}
\end{table}
On 2003 June 30 the {\em MOST} satellite was launched into a polar, Sun-synchronous circular orbit with an altitude of 820~km.
It carries a 15-cm Rumak-Maksutov telescope with a single, custom broadband (350 to 750 nm) optical filter  attached to a CCD photometer.
{\em MOST}'s orbital period is 101.413 minutes, which corresponds to an orbital frequency of $\sim$14.2~d$^{-1}$.

{\em MOST} observed HD~34282 from 2007 December~4 to 2008 January~4 (31 days) as an uninterrupted Direct Imaging Target.
Individual exposures were  $\sim 3$~s each with 20 consecutive images stacked on board the satellite, giving a sampling time of $\sim 60$~s per co-added measurement. 
The {\em MOST} on-board clocks are updated with time stamps from the ground stations (synchronized with atomic time) every day.  Individual exposure start times are accurate to 0.01\,s resulting in an even higher accuracy for the total exposure times. Barycentric corrections are for the Earth.

Two independent methods for the reduction of {\em MOST} Direct Imaging Photometry have been developed: 1) a combination of classical aperture photometry and point-spread function fitting to the Direct Imaging Subrasters, as developed by \citet{2006ApJ...646.1241R};  2) a data-reduction pipeline for space-based, open-field photometry that includes automated corrections for cosmic-ray hits and a stepwise pixel-to-pixel decorrelation of stray-light effects on the CCD
\citep{2008CoAst.152...77H}.
Tests on several {\em MOST} data sets in which both methods were used gave no significant differences in the quality of the extracted
light curves \citep[\eg see][]{2009A&A...494.1031Z}.
For HD~34282, the {\em MOST} Direct Imaging data were reduced using the method developed by \citet{2006ApJ...646.1241R}. 
The resulting light curve consists of 23093 data points, as outliers from the phases of the {\em MOST} orbit with the highest stray-light counts had to be discarded in the reduction.
Additionally, there are two gaps in the light curve which resulted from interruptions of the HD~34282 observations for a high-priority MOST target of opportunity.

HD~34282 has a bright companion at a distance of about 3 arcminutes.
In Direct Imaging Mode, the focal-plane scale is about 3 arcseconds per pixel and the raster used is about 20 pixels wide resulting in a mask size of about 60 arcseconds.
This safely excludes contamination of the HD~34282 MOST light curve by a brighter star about 180 arcseconds away.

\begin{figure}
\includegraphics[width=\columnwidth]{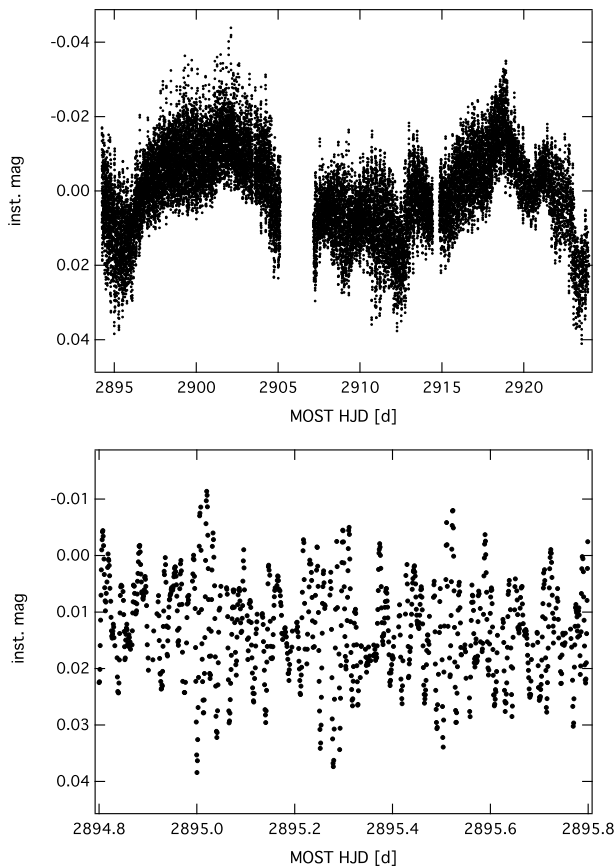}
\caption{Top panel: complete 31-day {\em MOST} light curve of HD~34282; bottom panel: one-day subset of the  {\em MOST}  light curve illustrating the pulsational variability.}
\label{lcs}
\end{figure}

\begin{figure*}
\includegraphics[width=1.80\columnwidth]{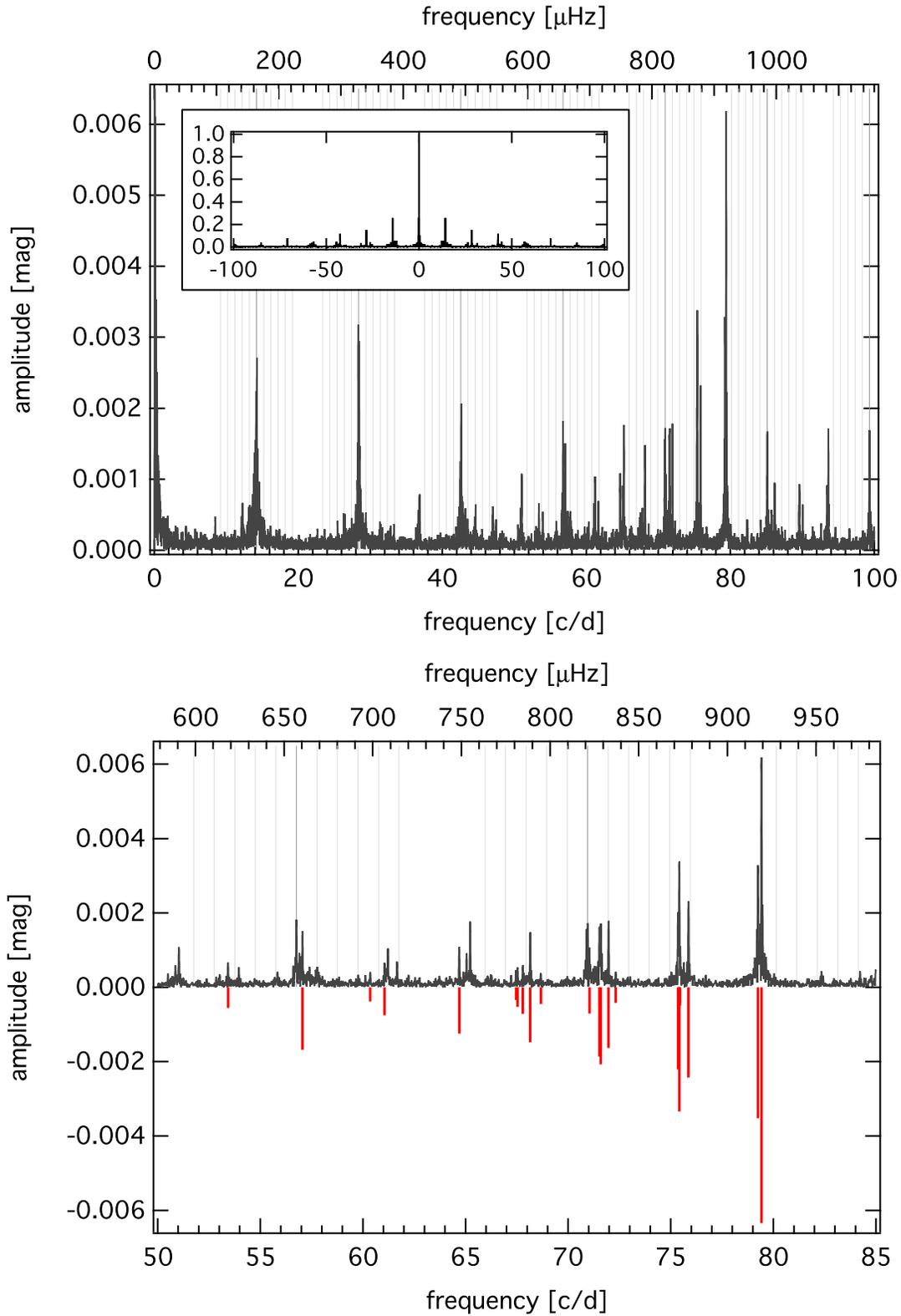}
\caption{Top panel: complete amplitude spectrum from 0 to 100~d$^{-1}$ (bottom axis) or from 0 to 1157.407~\mhz\, respectively (top axis),  where the corresponding spectral window is given as inset; bottom panel: zoom into the amplitude spectrum between 50 and 85~d$^{-1}$ (bottom axis) or from 578 to 984~\mhz\, respectively (top axis) where the 22 identified pulsation frequencies are identified (in red using negative values). Note that unmarked peaks are aliases of the pulsation frequencies with the {\em MOST} orbital frequency.
Dark grey lines mark the multiples of the {\em MOST} orbital frequency and light grey lines are the respective 1~d$^{-1}$ side lobes.}
\label{amps}
\end{figure*}

For the frequency analysis the {\sc Period04} \citep{2005CoAst.146...53L} and {\sc SigSpec} \citep{2007A&A...467.1353R} software packages were used, and the respective results compared for consistency. 
HD~34282 was observed during a period of relatively unvarying extinction, as only moderate peak-to-peak irregular variability in the integrated light  of less than 0.1 magnitude is observed (top panel of Fig.~\ref{lcs}).
22 frequencies thought to be intrinsic to the star (\eg non-instrumental)  are found between 50 and 85~d$^{-1}$ (578 and 926~$\mu$Hz), corresponding to periods ranging from 18 to 30 minutes.
These frequencies are listed in Table~\ref{tab:freqs} and shown in Fig.~\ref{amps}.
Ten of these frequencies were previously detected by \citet{2006MmSAI..77...97A}, given by the A06 ID in the final column of Table~\ref{tab:freqs}.
Twelve new, lower-amplitude frequencies are detected due to the enhanced sensitivity of {\em MOST} compared to the ground-based instrumentation.

A number of frequencies in the power spectrum of Fig.~\ref{amps} (\eg at 51~d$^{-1}$) are not intrinsic but are either identified as aliases of the pulsation frequencies with the {\em MOST} orbital frequency, and disappear with appropriate pre-whitening, or as instrumental frequencies related to the orbit of the satellite. 
The modulation of stray light with the orbital period of the satellite is itself modulated slightly with a 1~d$^{-1}$  frequency.  The Sun-synchronous orbit of MOST brings it over almost the same point on Earth after one day.  The albedo pattern of the Earth introduces a 1~d$^{-1}$ modulation of the amplitude of the 14.2~d$^{-1}$ modulation of
scattered Earthshine. Therefore, all significant peaks that can be related to these instrumental effects within the frequency resolution were discarded.

None of the 22 frequencies are identified as combination frequencies of the others, {\em i.e.} all would appear to be independent frequencies.

Collectively the frequency spectrum of HD~34282 in Fig.~\ref{amps} is quite striking, with distinct groups of frequencies spaced every 44~\mhz, ending abruptly at the high frequency end. 
Other stars that exhibit similar frequency groupings are 44 Tau \citep{2008A&A...478..855L} and HD~144277 \citep{2011A&A...533A.133Z}.
None of the stars for which frequency clustering has been previously observed shows the sudden cut-off in amplitude at high frequencies that HD~34282 does.

\section{Stellar and asteroseismic modelling.}
\subsection{HR Diagram position}
\label{sec:HRD_pos}
\begin{figure}  
\begin{centering}
\includegraphics[width=\columnwidth]{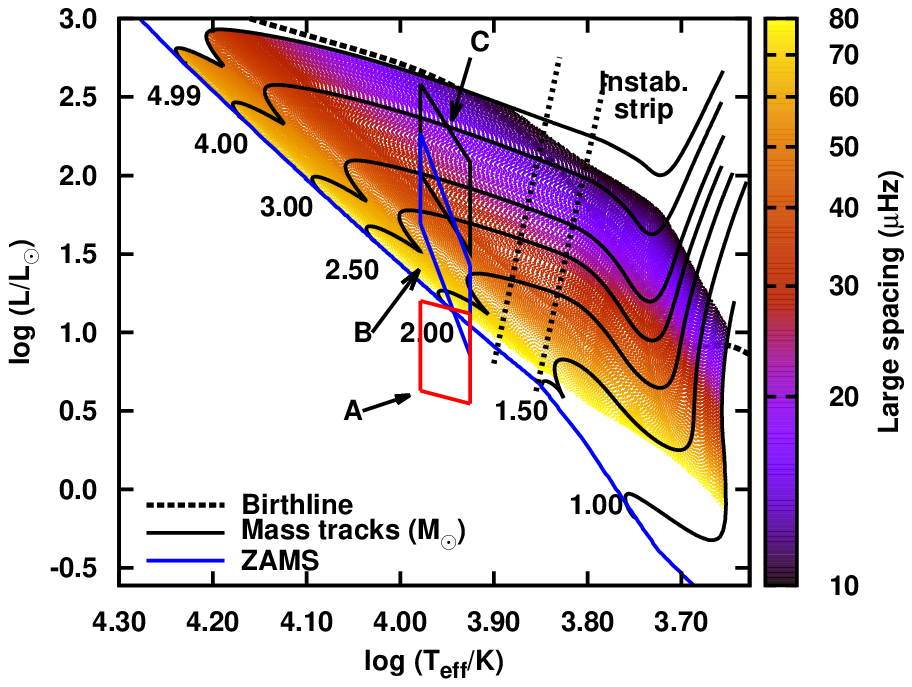}
\caption{Possible positions of HD~34282 in the HR diagram: ``A'' from \citet{2003AJ....126.2971V}; ``B'' and ``C'' from Tycho photometry of Fig.~\ref{fig:VT}.   The instability strip for the first three radial modes of PMS stars is as originally calculated by \citet{1998ApJ...507L.141M}.  The colouring indicates the large spacing of models that fall within the observational range of HD~34282.}
\label{fig:HRDpos}
\end{centering}
\end{figure}

Determining an HAe star's theoretical HR diagram position from observations is particularly difficult, mostly due to the circumstellar material obscuring the star.
Given these difficulties, we purposely take a broad approach to illustrate the relative precision of the (below) asteroseismic analysis.
A forthcoming paper by  Amado et al. (in preparation) will address the fundamental parameters of the star in much greater detail, parameters which will need to be reconciled with the asteroseismic analysis.
Here, three potential positions for HD~34282 are considered, displayed in Fig.~\ref{fig:HRDpos}, all derived using broad-band photometry.
An effective temperature  ($T_{eff}$) range of 8420 to 9520~K is used, encompassing spectral classes determined by various authors ranging from A0 to A3.
In the HR diagram this places the star blueward of the instability strip as determined by \citet{1998ApJ...507L.141M} for the first three ($n=0$ to 2) radial modes of PMS stars (driven by the $\kappa$-opacity mechanism).
However, HD~34282 (as will be shown) is pulsating in higher overtones than $n=2$, and so its placement to the left of the instability strip in the HR diagram is not surprising, and follows a trend outlined in \citet{2011PhDT......Casey}.
$T_{eff}$ and bolometric correction (BC) values were extracted from the spectral types by comparison to the tables published in \citet{1996imsa.book.....O}, based upon the work of  \citet{1982lbg6.conf.....schmidt}.
These same tables were used to determine the amount of reddening, $E(B-V)$,  by comparing the intrinsic ($B-V$)$_0$ colours to broad-band observations of $B-V$.
The overall extinction in $V$, $A_V$,  was determined, assuming
$A_V = R_V  E(B-V)$
where $R_V = 3.1$, and is the empirically-determined ratio of selective-to-overall extinction, as measured, {\em e.g.}, by \citet{1994RMxAA..29..163T} that usually applies to extinction caused by interstellar material.
Importantly, in the case of HAe stars that are often subject to large and variable levels of circumstellar extinction, this assumption is quite likely wrong.
At high levels of $A_V$ (more than 1.5 magnitudes or so), stars may start to appear bluer instead of redder with increased $A_V$, a result of the star's surrounding dust clouds reflecting light into the line of sight of the observer, causing a bluing effect.
In this case $R_V=3.1$ will underestimate the intrinsic brightness of the star, miss-positioning it in the HR diagram.  Unfortunately, without studies such as by \citet{1996A&A...309..809V}, the true value of $R_V$ cannot be determined.

This is illustrated by position A of Fig.~\ref{fig:HRDpos}, derived from $V=9.84 \pm 0.02$, $B-V = 0.17 \pm 0.02$ of \citet{2003AJ....126.2971V}. 
Given the range of possible spectral classes and distances (the latter from the Hipparcos range cited in the introduction), the resulting error box is a trapezoid in the HR diagram.
Significantly, position A falls essentially below the ZAMS, consistent with HD~34282 suffering from the bluing effect, but it is also possible that the true parallax falls outside the $1\sigma$ Hipparcos parallax uncertainty quoted in Section~\ref{sec:HD34282}, which could also result in an underestimate of the star's luminosity.

\begin{figure}
\begin{centering}
\includegraphics[width=0.75\columnwidth]{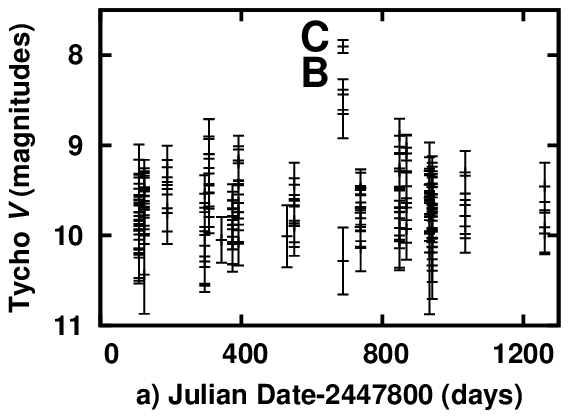}
\includegraphics[width=0.75\columnwidth]{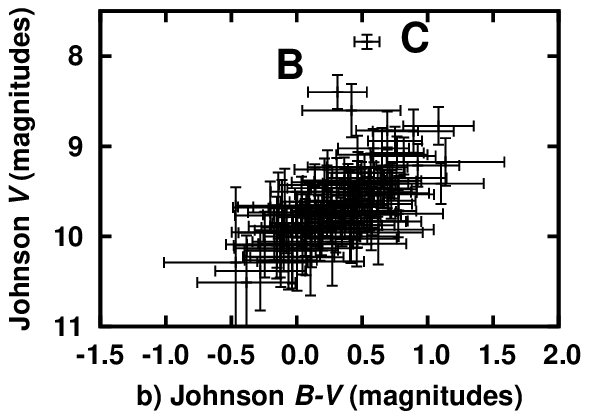}
\caption{a) Tycho $V$-band light curve for HD~34282; b)  CMD for HD~34282, each point representing a different point in the Tycho 2 light curve.  Johnson magnitudes were converted from Tycho magnitudes.  Positions B and C provide the corresponding positions in the HR diagram in Fig.~\ref{fig:HRDpos}.}
\label{fig:VT}
\end{centering}
\end{figure}

If the star {\em is} subject to blueing, this may be revealed by time-series broad-band photometry.
Positions B and C are derived from Tycho~2 photometry \citep{2000A&A...355L..27H}, displayed in Fig.~\ref{fig:VT}, in which the light curve of the star in Tycho $V$ ($V_T$) magnitudes are shown (Tycho $B$, $B_T$, also exists, but is not shown here).
$B_T$ and $V_T$ filters can be transformed to  Johnson $B$ and $V$ filters using the prescription detailed in Appendix C of \citet{2002AJ....124.1670M}, and addendum \citep{2006AJ....131.2360M}.
With this, a Johnson $V$, $B-V$ colour-magnitude diagram has been constructed in Fig.~\ref{fig:VT}b, displaying a potential blueing effect.
There is one distinct brightening event, during which the starlight is more likely to obey an $R_V=3.1$ reddening law, and hence two data points from this event, marked ``B'' and ``C'' in Fig.~\ref{fig:VT} (both parts), are used to calculate the corresponding HR diagram positions from Fig.~3.
Unfortunately, the data set is not complete enough to determine whether $R_V=3.1$ is correct for either point (or what the correct value should be), however corrections for these two points will certainly give better estimates of $A_V$ than the data used to determine position A.
Unsurprisingly, positions B and C are quite a bit brighter than position~A.
As will be shown below, position B seems to agree best with the asteroseismic analysis. 

The above analysis shows the broad range of intrinsic luminosities that can be obtained for an HAe star if the data are not treated properly. 
The above is not meant to be a final analysis, but an indicator that more observations are needed.
Long term multi-filter observations  are currently under way to determine the true reddening law for of HD~34282, plus a forthcoming paper will address the fundamental parameters of HD~34282 in much greater detail (Amado et al., in preparation).

\subsection{Asteroseismic analysis}
\label{sec:asteroseis}
\subsubsection{Large spacing}
The unique frequency spectrum of HD~34282 suggests that pulsation frequencies are being triggered around a ``main'' frequency, the central frequency of each group perhaps indicative of successive orders of radial pulsation.
If this is true, then the group spacing of approximately 44~\mhz~will match the asteroseismic average large spacing between radial orders of pulsation \citep{2010aste.book.....A}.
Figure~\ref{fig:HRDpos} shows the large spacings of the stellar models under consideration.
Models with a large spacing between 40 and 50~\mhz~that would match the observed group spacings of HD~34282 coincide with position ``B'' in the HR diagram.
Further analysis, presented below, strongly support this general region  as the true location of HD~34282 in the HR diagram.

The stellar models used here are the same as used in \citet{2009ApJ...704.1710G} for the PMS \DSS s in NGC~2264.
To summarize, a large grid of stellar models, closely spaced in position in the HR diagram was constructed using the {\sc yrec} stellar evolution code \citep{2008Ap&SS.316...31D}.
Solar metallicity ($Z=0.02$) models between 1.00 and 5.00~$\msun$ and low metallicity ($Z=0.004$) models between 1.00 and 3.00~$\msun$ (in increments of 0.01~$\msun$ for both cases) were considered.
Both PMS and post-ZAMS evolutionary tracks were evolved, the former starting on the Hayashi track with a polytrope, and ending at the ZAMS, and the latter using the ZAMS as a starting point, and ending near the base of the red giant branch.
For each model within an evolutionary track, adiabatic and non-adiabatic oscillation frequencies $\nu_{n{\ell}m}$ were calculated using Guenther's non-adiabatic stellar pulsation program \citep{1994ApJ...422..400G}.
Radial orders $n=0$ to 30 and azimuthal orders, $\ell=0$ to 3 were calculated.
The effects of rotation were not considered.

\subsubsection{``Averaged'' frequencies.}
\begin{figure}
   \begin{centering}
    	\includegraphics[width=\columnwidth]{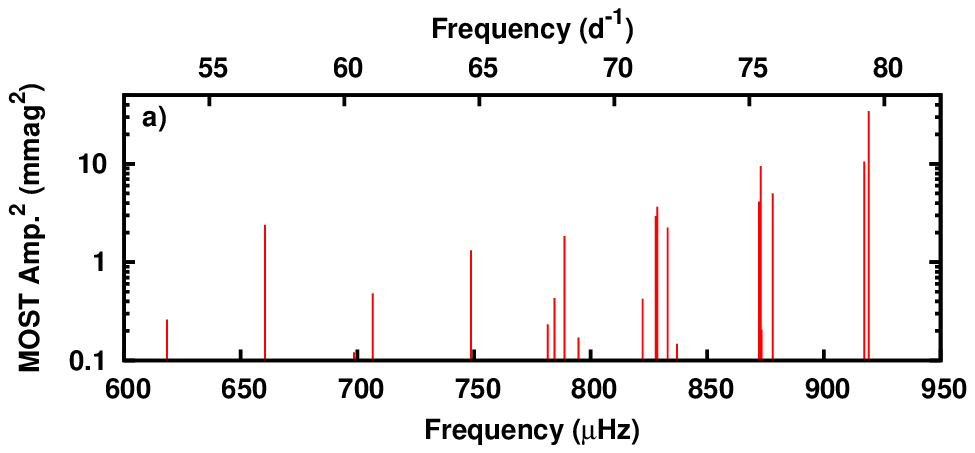}
    	\includegraphics[width=\columnwidth]{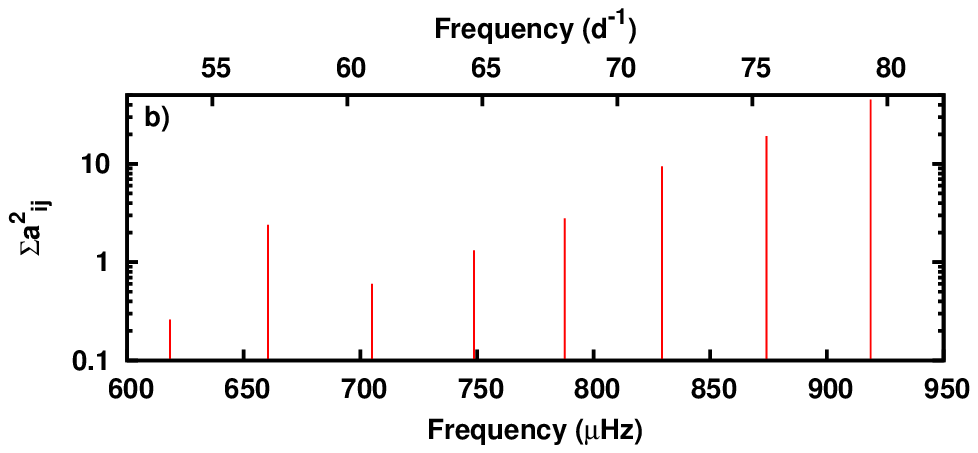}
    \caption{Squared amplitudes for a) unweighted and b) weighted frequencies of HD~34282 as detected by {\em MOST}.}
    \label{fig:V1366_Ori_weighted}
    \end{centering}
\end{figure}

It is possible that some of the individual frequencies within each group could be caused by time variations in the amplitude of, for example, the radial modes, including frequencies that might be damped.
In this case, a Fourier transform spectrum would show broadened or distinct modes depending on the resolution of the time-series data.
Note that in this case, not all of the detected 22 frequencies would be separate pulsation frequencies, and so caution must be taken when comparing the observed frequencies to models.
Unfortunately, for the closest pairs, the  temporal extent of the observations ($\sim$ 31~days) is insufficient to test this hypothesis, as the frequency resolution from considering only part of the  light curve is not high enough to distinguish between \eg $f_1$ and $f_2$.
However, if all the frequencies {\em are} stable then  some physical phenomenon must be selectively driving the pulsation frequencies in these distinct groups.
Here we postulate each group is triggered around successive radial orders of pulsation, and as an experiment  construct an ``average'' frequency for each group for comparison to models.
We computed the weighted average of each frequency group $H_j$, according to:
\begin{equation}
\label{eq:ampav}
H_j=\frac{\sum_{i=1}^{N_j}{f_{ji}a^2_{ji}}}{\sum_{i=1}^{N_j}a^2_{ji}},
\end{equation}
where $a_{ji}$ and $f_{ji}$ are the $i^{th}$ constituent amplitudes and frequencies of the $j^{th}$ grouping, with resultant weighted frequency, $H_j$.  $N_j$ is the number of frequencies included in the $j_{th}$ group.
The weighted averaged frequencies for each group are listed in Table~\ref{tab:V1366_Ori_weighted} in order of increasing frequency.
The squared amplitude for each $H_j$ is given by the sum of the constituent squared amplitudes.
For comparison, Fig.~\ref{fig:V1366_Ori_weighted} shows the a) unweighted and b) weighted frequency spectra of the star.
\begin{table}
    	\begin{minipage}{\columnwidth}
           \resizebox{\columnwidth}{!}
   	{
           	\begin{tabular}{lcrrrrrrr}
 \hline
$H_j$  &   Orig.    &  $j,i$ & $f_{ji}$ & $f_{ji}$  & $a_{ji}$ & $a_{ji}^2$   & $\sum a^2_{ji}$ &$H_{j}$  \\
ID     &   ID       &        & ($\mu$Hz)&(d$^{-1}$) & (rel.amp)& (rel.amp$^2$)& (rel.amp$^2$)   &($\mu$Hz)\\
\hline
$H_1$  & $f_{16}$   &  1,1   &   618.37 & 53.427    & 0.510    &  0.260       &  0.260          & 618.4   \\
       &            &        &          &           &          &              &                 &         \\
$H_2$  & $f_8$      & 2,1    &  660.42  & 57.060    & 1.549    &  2.400       & 2.400           &  660.4  \\
       &            &        &          &           &          &              &                 &         \\
$H_3$  & $f_{21}$   & 3,1    &   698.53 & 60.353    & 0.348    &  0.121       & 0.604           & 705.0   \\
       &  $f_{12}$  & 3,2    &  706.63  &61.053     & 0.695    &   0.483      &                 &         \\
       &            &        &          &           &          &              &                 &         \\
$H_4$  & $f_{11}$   & 4,1    & 748.79   &64.695     & 1.148    & 1.317        & 1.317           & 748.8   \\
       &            &        &          &           &          &              &                 &         \\
$H_5$  & $f_{22}$   &  5,1   & 780.84   &67.465     & 0.317    & 0.101        & 2.786           & 787.6   \\
       & $f_{17}$   &  5,2   & 781.65   & 67.534    & 0.483    & 0.233        &                 &         \\
       & $f_{13}$   &  5,3   & 784.58   & 67.787    & 0.658    & 0.433        &                 &         \\
       & $f_{10}$   & 5,4    & 788.80   & 68.152    & 1.360    & 1.849        &                 &         \\
       & $f_{19}$   & 5,5    & 794.78   & 68.669    & 0.413    & 0.171        &                 &         \\ 
       &            &        &          &           &          &              &                 &         \\
$H_6$  & $f_{14}$   & 6,1    & 822.26   &71.043     &  0.651   & 0.424        & 9.434           & 829.3   \\
       & $f_7$      & 6,2    & 827.84   & 71.525    & 1.716    &   2.946      &                 &         \\
       & $f_6$      & 6,3    & 828.58   & 71.589    & 1.913    &    3.658     &                 &         \\
       & $f_9$      & 6,4    & 833.01   & 71.972    & 1.502    &   2.257      &                 &         \\
       & $f_{20}$   & 6,5    & 837.03   & 72.319    & 0.385    &    0.148     &                 &         \\
       &            &        &          &           &          &              &                 &         \\
$H_7$  & $f_5$      & 7,1    & 872.18   & 75.356    & 2.033    & 4.135        &   19.198        & 874.1   \\
       & $f_{15}$   & 7,2    & 872.70   & 75.402    & 0.606    & 0.368        &                 &         \\
       & $f_3$      & 7,3    & 872.87   & 75.416    & 3.080    &  9.485       &                 &         \\
       & $f_{18}$   & 7,4    & 873.24   & 75.448    & 0.453    &  0.205       &                 &         \\ 
       & $f_4$      & 7,5    & 878.05   & 75.864    & 2.237    &   5.006      &                 &         \\
       &            &        &          &           &          &              &                 &         \\
$H_8$  & $f_2$      & 8,1    & 917.27   & 79.252    & 3.250    & 10.561       & 44.903          & 918.8   \\
       & $f_1$      & 8,2    & 919.24   & 79.423    & 5.860    &  34.342      &                 &         \\
\hline
\end{tabular}

          	}
    	 \end{minipage}
    \caption{Weighted frequencies for HD~34282.}

    \label{tab:V1366_Ori_weighted}
\end{table}

In order to compare these values to models, uncertainties need to be assigned to the  $H_i$, a somewhat speculative venture.
We choose an uncertainty of $\pm 1$~\mhz~for each $H_j$, matching the approximate range of the two constituent frequencies of the highest-amplitude group, $H_8$.

To locate the best match between the observed pulsation spectrum and the model spectra we quantified the fits using the $\chi^2$ equation:
\begin{equation}
\label{eq:chisq}
\chi^2= \frac{1}{N}\sum^{N}_{i=1}\frac{(\nu_{obs,i}-\nu_{mod,i})^2}{\sigma^2_{obs,i}+\sigma^2_{mod,i}},
\end{equation}
where $N$ is the number of observed frequencies, $\nu_{obs,i}$ and $\nu_{mod,i}$ are the $i^{th}$ observed and model frequencies respectively, and $\sigma^2_{obs,i}$ and $\sigma^2_{mod,i}$ are the $i^{th}$ observed- and model-frequency uncertainties respectively.
As in \citet{2009ApJ...704.1710G}, $\sigma^2_{mod,i}$ is small compared to $\sigma^2_{obs,i}$ and is ignored for the purposes of these calculations.

\begin{figure*}
\begin{center}
  \begin{minipage}[l]{\columnwidth}
           	\includegraphics[width=0.75\textwidth]{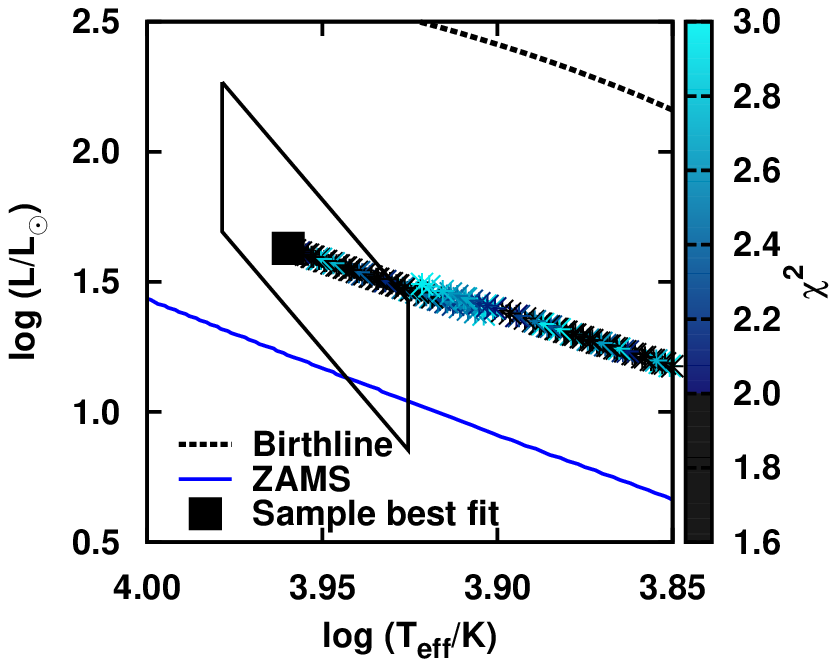}
	\includegraphics[width=0.75\textwidth]{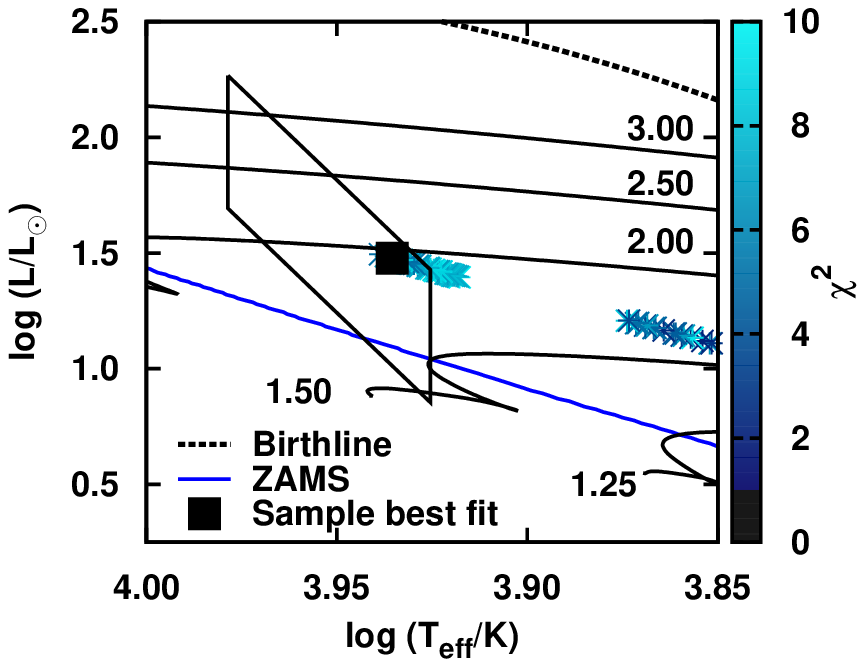}
  \end{minipage}
 \begin{minipage}[r]{0.98\columnwidth}
    	\includegraphics[width=0.75\textwidth]{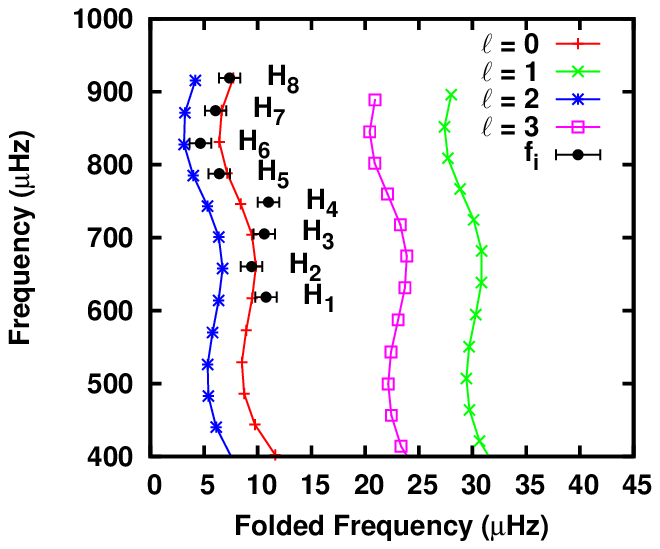}
	 \includegraphics[width=0.75\textwidth]{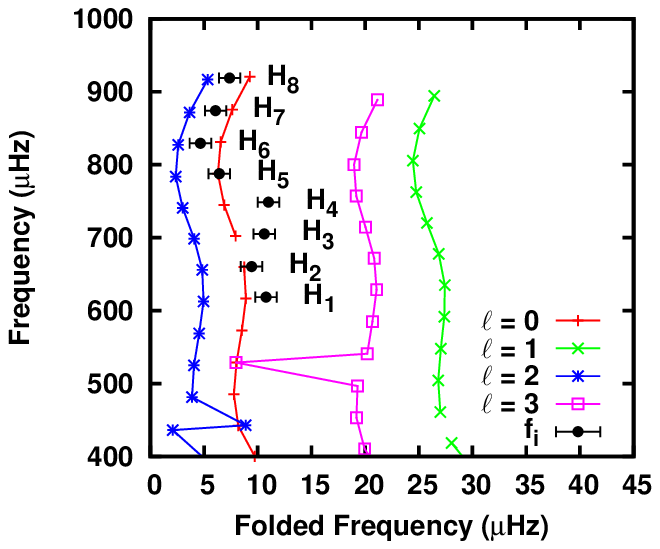}
\end{minipage}
 \caption{\chisq fits to weighted frequencies of HD~34282, $\ell = 0$ modes only.  Top Row:  $Z$=0.02, best model fit \chisq=1.7, $M$=2.30~$\msun$, \logT=3.960, \logl=1.629.
 Bottom row: $Z$=0.004, best model fit \chisq=4.7, $M$=1.92~$\msun$, \logT=3.935, \logl=1.478.  Both echelle diagrams use a folding frequency of 43.4~\mhz.}
\label{fig:V1366_Ori_z02_echelle}
\end{center}
\end{figure*}

The top two panels of Fig.~\ref{fig:V1366_Ori_z02_echelle} show \chisq fits to our $Z=0.02$ PMS model grid, along with a sample echelle diagram.
Only models within the grid that have \chisq $< 3.0$ are shown.
The trapezoid corresponds to position~B for HD~34282, and the large black square to our model best fit to the weighted-averaged frequencies.
The fits are to radial-order modes only. 

The bottom two panels show the fits to a $Z=0.004$ model grid.
The best fit to the $Z=0.02$ is slightly better than the best fit to the lower metallicity,  and thus marginally favours a $\lambda$ Bootis nature for the star.
The results are encouraging particularly given the speculative nature of the analysis.
In both cases the observed spectrum of HD~34282 appears to correspond to high radial-order modes, with $n=13$ to 20.
Of particular note, in Fig.~\ref{fig:V1366_Ori_z02_echelle} there are no theoretical frequencies displayed above $n=20$, as any pulsation frequency above this value would be greater than the model's acoustic cut-off frequency, a theoretical maximum pulsation frequency for the model, the consequences of which will be addressed in the following subsection.

\begin{figure}
\begin{centering}
\includegraphics[width=\columnwidth]{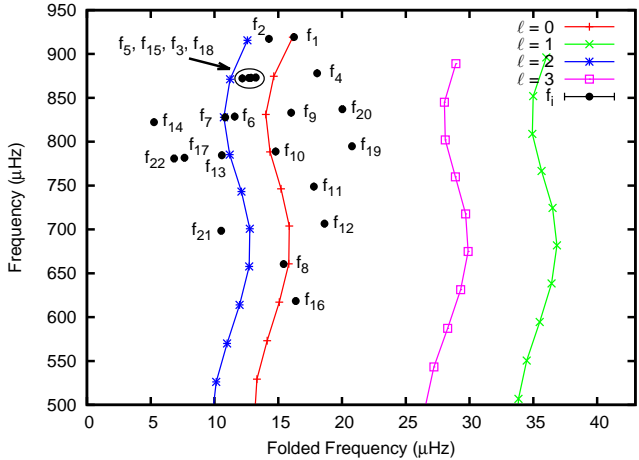}
\caption{Echelle diagram of the unweighted frequencies of HD~34282 compared to the $Z=0.02$ model of Fig.~\ref{fig:V1366_Ori_z02_echelle}.  The frequencies do not extend to $\ell=1$ nor 3 modes.}
\label{fig:V1366_echelle}
\end{centering}
\end{figure}

Note that if we plot the original frequencies in a similar echelle diagram as in the top of Fig.~\ref{fig:V1366_Ori_z02_echelle} the frequencies would scatter slightly about the $\ell=0$ (and $\ell=2$) model frequencies, but would not spread to the $\ell=1$ (and 3) model frequencies (see Fig.~\ref{fig:V1366_echelle}).

Further \chisq~fits were performed with $\ell = 1$ modes only, but the results are not shown.
The results are similar in quality, with a line of good fits that are a bit cooler than those of the $\ell = 0$ case and return models with similar large spacings and  radial orders that run from $n=13$ to 20.
Simultaneous fits to both $\ell = 0$ and 1 modes did not yield good results, indicating that the frequency groups are not probing the separation between $\ell = 0$ and 1 modes.
Similarly, fits of the unweighted frequencies to $\ell=0$ to 3 modes did not yield good results, yielding extremely high \chisq values of around $10^5$.
Regardless, overall, the frequency groups displayed in the pulsation spectrum of HD~34282 are consistent with those groups representing successive order of radial pulsation of the star, the mechanism for this pattern unknown at this time.

\subsubsection{Acoustic cut-off frequency}

The average amplitudes of the frequencies of the groups increases monotonically with frequency from 700~\mhz\, onward, then abruptly stops with no periodic signal detected above $\sim 920$~\mhz. We believe that this abrupt drop off corresponds to the acoustic cut-off frequency for HD~34282 and that models with acoustic cut-off frequencies above this frequency can be ruled out.  

In standard theory, p-modes above the acoustic cut-off frequency are no longer reflected back at the surface but continue on as travelling waves into the atmosphere of star where they quickly radiate away their energy.
Indeed, in the early days of helioseismology we expected to see an abrupt drop off in the Sun's p-mode spectrum above the theoretically-predicted solar acoustic cut-off frequency \citep{1992A&A...266..532F}.
In fact, in the case of the Sun, regularly spaced modes above the acoustic cut-off frequency are observed with amplitudes that decrease at higher frequencies \citep{1988ESASP.286..279J,1988ApJ...334..510L}.
Again we expected these modes to form a continuous spectrum, being stochastically driven travelling waves.
But in the case of the Sun the modes are spaced out in a pattern that mimics the spacing of the trapped p-modes below the acoustic cut-off frequency.
Currently, we believe that the pseudomodes, as they are called, are driven by turbulent convection in the atmosphere and that their regular spacings are caused by simple geometric interference as the waves travel around the star \citep{1994ApJ...428..827K,1998ApJ...504L..51G,2011ApJ...743...99J}. 

For most stars, the amplitudes of the p-modes do decrease below the detection threshold well before reaching the theoretical acoustic cut-off frequency.
Therefore, the rise in amplitudes with a sudden drop off at the highest observed frequency for HD~34282 is unusual. 

\begin{figure}
\begin{centering}
\includegraphics[width=\columnwidth]{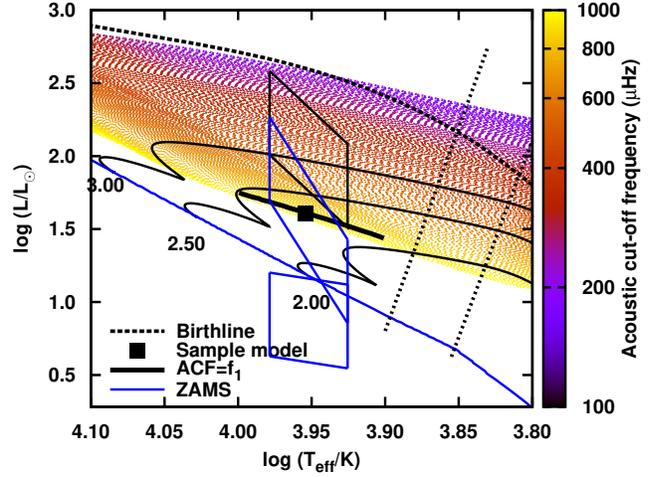}
\caption{Acoustic cut-off frequencies of the PMS models. 
Models with an acoustic cut-off frequency less than 100~\mhz~and greater than 1000~\mhz~are not shown, as they are out of observed \DS~pulsation range.
Only models between the thick solid line and the ZAMS can support the pulsations of HD~34282.  The black square indicates the $Z=0.02$ sample model of Fig.~\ref{fig:V1366_Ori_z02_echelle}.}
\label{fig:PMSacf}
\end{centering}
\end{figure}

At this time we do not believe that the acoustic cut-off frequency is below the highest observed frequency in HD~34282, that is, we do not believe any of the observed modes are pseudomodes.
If the higher frequency modes we observe were pseudomodes then they should have short life times and random phases. Indeed this could provide a possible explanation for the multiple peaks within each group.
Furthermore, as Fig.~\ref{fig:PMSacf}, shows, within the uncertainties of HD~34282's  HR-diagram position, viable models with acoustic cut-off frequencies as low as $\sim 300$~\mhz~are possible.
But what leads us to doubt this possibility is the fact that the amplitudes of the averaged modes continue to increase with frequency.
All currently proposed models to explain the existence of the Sun's pseudomodes predict a drop in amplitudes with increasing frequency. 

Therefore, we speculate that the acoustic cut-off frequency, indeed, provides an upper frequency limit to the regularly spaced p-modes in HD~34282 and that the acoustic cut-off frequency is near or above $\sim~920$~\mhz.

Fig.~\ref{fig:PMSacf} shows the acoustic cut-off frequencies for PMS $Z=0.02$ model-grid stars.
Under our assumption, we can eliminate models whose acoustic cut-off frequencies are lower than the highest frequency observed, i.e., $f_1$ at $\sim 920$~\mhz.
As shown in Fig.~\ref{fig:PMSacf}, for fixed $T_{eff}$, the acoustic cut-off frequency decreases as the luminosity of the model star increases.
Hence, for HD~38282 there is a near-horizontal line in the HR diagram (solid, black line passing through the sample model in Fig.~\ref{fig:PMSacf}), above which models have too low an acoustic cut-off frequency to meet our assumption (that the acoustic cut-off frequency is above 920~\mhz).
Only the models in the region between this line and the ZAMS are viable. 

Finally we note that our best model fit to the averaged mode frequencies (see Fig.~\ref{fig:V1366_Ori_z02_echelle}) lies just below this line, in agreement with our assumption that the highest observed frequency is just below the acoustic cut-off frequency.

\subsubsection{Rotational splitting}
\begin{figure}
\begin{centering}
\includegraphics[width=\columnwidth]{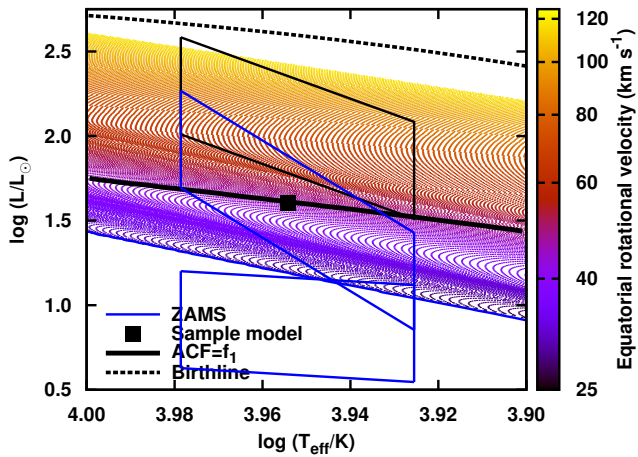}
\caption{Surface equatorial rotational velocities required to give $\Delta f = 4$~\mhz~(colour bar).
Models between the solid black line and the ZAMS are those with an acoustic cut-off frequency greater than  $f_1$ = 919.24~\mhz.}
\label{fig:V1366_Ori_HR_maxrot}
\end{centering}
\end{figure}

Here we comment upon the minimum role that rotation must play in the pulsation spectrum of the star.
Individual (unaveraged) frequencies could not be identified as distinct low $\ell$-valued modes, so we consider the possibility that the frequencies within groups correspond to a rotationally-split mode.
Within a group some frequencies have separations of about 2 to 4~\mhz, but
given the observed \vsini, Fig.~\ref{fig:V1366_Ori_HR_maxrot} shows these frequencies  are unlikely to be from the same $\ell$ mode.
A first-order estimate of the rotational splitting, $\Delta f$, between two successive $m$ modes (\eg between $m$ and $m+1$) within a multiplet implied by $v$ (the surface equatorial velocity) is  $\Delta f = 1/{\Omega}$, where $\Omega$ is the surface equatorial rotation period of the star, $\Omega = (2\pi R_*)/v$, and $R_*$ is the radius of the particular stellar model in question \citep{2010aste.book.....A}.\footnote{It is important to note that for large values of the ratio $R_*/v$, $\Delta f_{min}$ does not apply in detail, but only on average
\citep[][Deupree, private communication]{2008ApJ...679.1499L,2010ApJ...721.1900D}.}
For  each model in the grid, Fig.~\ref{fig:V1366_Ori_HR_maxrot} shows the equatorial velocity required to give $\Delta f = 4$~\mhz.
Only the most luminous models have $v$ consistent with the \vsini~$=~129 \pm 11$~km~s$^{-1}$ of \citet{2001A&A...378..116M}.
The solid black line that parallels the ZAMS in Fig.~\ref{fig:V1366_Ori_HR_maxrot} separates models that are consistent with the acoustic cut-off frequency constraint (below the line) from models that do not (above the line).
If we accept the acoustic cut-off frequency constraint, then rotation as a possible source of the 2 to 4~\mhz~differences between frequencies within each group is ruled out.
If rotational splittings are present in the spectrum then the splittings must be at least $\sim 10$~\mhz.
As rotation rates become higher, rotational splittings become progressively non-linear, multiplets originating from different unsplit modes begin to overlap, and mode identification becomes more difficult.
Detailed 2D calculations, such as those by \citet{2010ApJ...721.1900D} are required to calculate these frequencies, and are beyond the current scope of this paper.

\section{Summary and conclusions}
\label{sec:conclusions}

{\em MOST} observations have discovered in the light curve of HD~34282 some 22 frequencies (12 more than previously observed), in which asteroseismic analysis suggests that these frequencies cluster around eight successive radial orders of pulsation.
The amplitude of each pulsation group grows with frequency, with an abrupt cut-off in power after the highest frequency detected.
We believe that the observed frequencies run right up to the acoustic cut-off frequency.
We think it is unlikely that we are observing pseudomodes, i.e. untrapped travelling waves, because the amplitude of the frequencies do not decrease with increasing frequency.
The average frequencies compiled from the groups of frequencies simultaneously fit the highest eight radial orders below the acoustic cut-off frequency ($n=13$ through 20 orders), and predict the acoustic cut-off frequency.

Although we have focused our discussion on fitting radial modes to the weighted-average frequencies, equally viable fits for $\ell=1$ p-modes, exclusively, can be achieved.
The $\ell=1$ best-fit models have similar large spacings, with only slightly lower temperatures and luminosities compared to the radial-mode models.
Therefore, regardless of which azimuthal order is considered, the best-fit models to the averaged frequencies occupy nearly the same position in the HR diagram as those shown in Fig.~\ref{fig:V1366_Ori_z02_echelle}.
The range of viable models in the HR diagram is also little affected by metallicity, with the lower-$Z$ example we tested predicting a slightly lower mass than the solar-$Z$ case.

The HR diagram position of HD~34282 remains difficult to determine -- the asteroseismic analysis in the work suggests position ``B'' in Fig.~\ref{fig:V1366_Ori_weighted} is the most appropriate, and that the star is therefore suffering from the blueing effect common to some heavily-obscured Herbig Ae stars.
Further work in this area is needed, and multi-filter observations of HD~34282 are currently under way to determine the true level of obscuration and reddening of the star.

The ultimate cause of the frequency groups is unknown, however it may be an example of mode trapping combined with large rotational splittings.
The large non-linear splittings expected with  \vsini$=~129 \pm 11$~km~s$^{-1}$, coupled with a selection mechanism that drives only  modes close in frequency space to that of a radial mode would explain the strange pattern observed in HD~34282.
Future theoretical calculations are needed to investigate this possibility.

\section*{Acknowledgements}
We wish to thank the referee for useful comments that allowed us to clarify certain parts of this paper.

KZ is a recipient of an APART fellowship of the Austrian Academy of Sciences at the Institute of Astronomy of the University Vienna.
DBG, MPC, SMR and AFJM acknowledge the funding support of the Natural Sciences and
Engineering Research Council of Canada.
AFJM also acknowledges the funding support of FQRNT.
RK and WWW are supported by the Austrian Science Fund (P22691-N16) and by the Austrian Research Promotion Agency-ALR.
PJA acknowledges financial support of the previous Spanish Ministry of Science and Innovation (MICINN), currently Ministry of Economy and Competitiveness, grant AYA2010-14840.
DD and ER acknowledge the support by the Junta de Andaluc\'{i}a and the Direcci\'{o}n General de Investigaci\'{o}n (DGI), project AYA2009-10394.
\bibliographystyle{mn2e}
\bibliography{HD34282}

\label{lastpage}

\end{document}